\documentclass[journal]{IEEEtran}
\usepackage{xcolor}
\usepackage{amssymb,amsthm,amsmath}

\newtheorem{theorem}{Theorem}[section]
\newtheorem{proposition}[theorem]{Proposition}
\newtheorem{lemma}[theorem]{Lemma}
\newtheorem{corollary}[theorem]{Corollary}
\newtheorem{definition}[theorem]{Definition}
\newtheorem{remark}[theorem]{Remark}
\newtheorem{example}[theorem]{Example}
\newtheorem{problem}{Open problem}

\begin{document}

\title{An inductive construction of minimal codes}

\author{Daniele~Bartoli, Matteo~Bonini, and Bur\c{c}in G\"une\c{s} 
	\thanks{D. Bartoli is with the Department of Mathematics and Computer Science, University of Perugia, Perugia, via Vanvitelli 1, 06123 Italy e-mail: daniele.bartoli@unipg.it}
	\thanks{M. Bonini is with the Department of Mathematics, University of Trento, via Sommarive 14, 38123 Italy e-mail: matteo.bonini@unitn.it}
	\thanks{B. G\"une\c{s} is with the Faculty of Engineering and Natural Sciences, Sabanci University, Tuzla, Istanbul, Turkey e-mail: bgunes@sabanciuniv.edu}
}

\maketitle
\begin{abstract}
We provide new families of minimal codes in any characteristic. Also, an inductive construction of minimal codes is presented.
\end{abstract}


\IEEEpeerreviewmaketitle

\section{Introduction}
\IEEEPARstart{L}{et} $\mathcal{C}$ be a linear code. A codeword $ c \in \mathcal{C}$ is called \emph{minimal} if its support (i.e., the set of nonzero coordinates of $ c $) does not contain the support of any other independent codeword. That is, $c$ is determined by its support up to a scalar factor.  

Minimal codewords can be used to describe access structures in linear code-based secret sharing schemes (SSS), see \cite{Massey1993,Massey1995}. 
A secret sharing scheme is a method to distribute shares of a secret to each of the participants $ \mathcal{P} $ in such a way that only the authorized subsets of $ \mathcal{P} $ (access structure $\Gamma$) could reconstruct the secret, see \cite{Shamir1979,Blakley1979}. 
A set of participants $A$ is said to be a \emph{minimal authorized subset} if $A\in \Gamma$ and no proper subset of $A$ belongs to $\Gamma$.

In  \cite{Massey1993,Massey1995} Massey considered the use of linear codes for realizing a perfect (i.e. all authorized sets of participants can recover the secret while unauthorized sets of participants cannot determine any shares of the secret) and ideal (i.e. the shares of all participants are of the same size as that of the secret) SSS. It turns out that the access structure of the secret-sharing scheme corresponding to an $[n,k]_q$-code $\mathcal{C}$ is specified by the support of minimal codewords in $\mathcal{C}^{\bot}$ having $1$ as the first component.

On one hand, minimal codewords are useful for constructing SSS, but on the other, given an arbitrary linear code $\mathcal{C}$ determining the set of its minimal codewords is a challenging task, see \cite{BMeT1978,BN1990}. 
Therefore, obtaining the access structures of SSS that are based on general linear codes is also difficult and it has been achieved only for specific classes of linear codes. 
This lead to the study of linear codes for which every codeword is minimal; see for instance \cite{CCP2014,SL2012}.

These problems gave rise to a new type of linear code, which has been introduced and investigated in order to utilize linear codes in SSS. A linear code is said to be \emph{minimal} if all its nonzero codewords are minimal. Most of the known families of minimal codes are in characteristic two. Due to their application to secret sharing schemes, the study of minimal codes has drawn immense attention in recent years. 

 A useful criterion for a linear code to be minimal is given by Ashikhmin and Barg in \cite{AB1998}.
\begin{lemma}\label{Th:Ashikhmin-Barg}
A linear code $\mathcal{C}$ over $\mathbb{F}_q$ is minimal if 
\begin{equation}\label{Eq:Ashikhmin-Barg}
\frac{w_{min}}{w_{max}} > \frac{q-1}{q},
\end{equation}
where $w_{min}$ and $w_{max}$ denote the minimum and maximum nonzero Hamming weights in $\mathcal{C}$,  respectively.
\end{lemma}

Families of minimal linear codes satisfying Condition \eqref{Eq:Ashikhmin-Barg} have been considered in several papers, e.g. see \cite{CDY2005,Ding2015,DLLZ2016,YD2006}. However, Condition~\eqref{Eq:Ashikhmin-Barg} does not necessarily hold for each minimal code. To this end, examples of minimal codes violating Condition~\eqref{Eq:Ashikhmin-Barg} have been constructed in \cite{CMP2013}, whereas in \cite{CH2017} the first infinite family of minimal binary codes has been presented by means of Boolean functions arising from simplicial complexes. Families of minimal binary and ternary codes have been constructed; see \cite{HDZ2018,DHZ2018}. Moreover, in \cite{BB2019} the authors provided examples of minimal codes for each field of odd characteristic. After that, in \cite{BB201?} minimal codes arising from cutting blocking sets were presented.

In this paper we construct examples of minimal linear codes in any characteristic; see Section \ref{Sec:Construction}. An interesting problem is to provide constructions of new minimal linear codes from old ones. In this direction, an example is given by the following result; see \cite[Proposition 5]{CMP2013}. 
\begin{proposition}  The product $\mathcal{C}_1\otimes \mathcal{C}_2$ of a minimal $[n_1, k_1, d_1]_q$-code $\mathcal{C}_1$ and of a
minimal $[n_2, k_2, d_2]_q$-code $\mathcal{C}_2$ is a minimal $[n_1 \times n_2, k_1 \times k_2, d_1 \times d_2]_q$-code.
\end{proposition}

In this paper we also provide inductive construction of minimal linear codes which satisfy an extra condition. 

The weight distribution of
a code allows the computation of the error probability of error detection and
correction with respect to some error detection and error correction algorithms; see \cite{Klove} for more details.

\section{Minimal codes and Secret Sharing Schemes}
Let $\mathcal{C}$ be an $[n,k]_q$-code, that is a $k$-dimensional linear subspace of $\mathbb{F}_q^n$. The support $Supp(c)$ of a codeword $c=(c_1,\ldots,c_n)\in \mathcal{C}$ is the set $\{i \in \{1,\ldots,n\} \ : \ c_i \neq 0\}$. Clearly, the Hamming weight $w(c)$ equals $|Supp(c)|$ for any codeword $c\in \mathcal{C}$. 
\begin{definition}\cite{Massey1993} A codeword $c\in \mathcal{C}$ is \emph{minimal} if it only covers the codewords $\lambda c$, with $\lambda \in \mathbb{F}_q^*$, that is 
p$$\forall \ c^{\prime}\in \mathcal{C} \Longrightarrow  \left(Supp(c) \subset Supp(c^{\prime}) \Longrightarrow \exists \lambda \in \mathbb{F}_q \ : \ c^{\prime}=\lambda c\right).$$
\end{definition}

\begin{definition}\cite{DY2003}
The code $\mathcal{C}$ is \emph{minimal} if every nonzero codeword $c \in \mathcal{C}$ is minimal.
\end{definition}

Let $G \in \mathbb{F}_q^{k\times n}$ be the generator matrix of $\mathcal{C}$ with columns $G_1,\ldots,G_n$ and suppose that no $G_i$ is the $0$-vector. The code $\mathcal{C}$ can be used to construct secret sharing schemes in the following way. The secret is an element of $\mathbb{F}_q$ and the set of participants  $\mathcal{P}=\{P_2,\ldots,P_{n}\}$. The dealer chooses randomly $u = (u_1, \ldots, u_{k}) \in \mathbb{F}_q^k$ such that  $s = u\cdot G_1$ and  computes the corresponding codeword $v=(v_1,\ldots,v_n)= u G$. Each participant $P_i$, $i\geq 2$,  receives the share $v_i$. A set of participants $\{P_{i_1},\ldots,P_{i_\ell}\}$ determines the secret if and only if $G_1$ is a linear combination of $G_{i_1},\ldots,G_{i_{\ell}}$; see \cite{Massey1993}. There is a one-to-one correspondence between minimal authorized subsets and the set of minimal codewords of the dual code $\mathcal{C}^{\bot}$.

\section{First construction}\label{Sec:Construction}

Let $q$ be prime power. Fix an integer $t\geq 2$ and consider the following matrix.

\begin{equation}\label{Eq:A}
A_{t,q}=
\left(
\begin{array}{ccccc| ccccc}
&&&&&&&&&\\
&&&&&&&&&\\
&&\mathbb{I}_t&&&&&\mathbb{B}_{t,q}&&\\
&&&&&&&&&\\
&&&&&&&&&\\
\end{array}
\right),
\end{equation}
where $\mathbb{I}_t$ is the identity matrix and $\mathbb{B}_{t,q} \in \mathcal{M}_{t\times \binom{t}{2}(q-1)}(\mathbb{F}_q)$ is a matrix whose columns are $e_i+\lambda e_j$, where $1\leq i<j\leq t$ and $\lambda\in \mathbb{F}_q^*$.

\begin{proposition}\label{Prop:Parameters}
Let $q\ge t-2$. Then the linear code $\mathcal{C}$ whose generator matrix is $A_{t,q}$ is a $\left[ \binom{t}{2}(q-1)+t,t,(t-1)(q-1)+1\right]_q$-code.
\end{proposition}
\proof
Clearly, the length is $\binom{t}{2}(q-1)+t$ and the dimension is $t$. 
Let us consider a codeword $\omega$ which is a linear combination of  $1\leq s\leq t$ rows, say $r_{i_1},\ldots,r_{i_s}$, of $A_{t,q}$. 
Without loss of generality we may assume that  $\omega$ has exactly $s$ nonzero entries among the first $t$.  

Suppose that $i< j$ and $i,j \in \{i_1,\ldots,i_s\}$. 
Thus, among the $(q-1)$ entries corresponding to columns $e_i+\lambda e_j$ exactly $q-2$ are nonvanishing. 

Consider now an index $i \in  \{i_1,\ldots,i_s\}$ and an index $j\notin \{i_1,\ldots,i_s\}$. 
Therefore, all the  $(q-1)$ entries corresponding to columns $e_i+\lambda e_j$ (or $e_j+\lambda e_i$) are nonzero. Summing up, the weight of $\omega$ is

$$w(\omega)=w_s=s+\binom{s}{2}(q-2)+s(t-s)(q-1).$$

By direct computations we have that 
\[
w_s-w_{s-1}=-t+(t-s)q+2.
\]
Then the minimum and the maximum are taken for $s=1$ and $s=t-1$. 

Thus,
\begin{eqnarray*}
w_{min}&=&1+(t-1)(q-1),\\
w_{max }&=&t-1+\binom{t-1}{2}(q-2)+(t-1)(q-1).
\end{eqnarray*}
Also, the codewords of weight $w_s$ are exactly $(q-1)^s$. 
\endproof

\begin{proposition}\label{Prop:Minimality}
The linear code $\mathcal{C}$ whose generator matrix is $A_{t,q}$ is a minimal code for which $w_{min}/w_{max }<(q-1)/q$.
\end{proposition}
\proof
We already saw in Proposition \ref{Prop:Parameters} that minimum and maximum weight are 
\begin{eqnarray*}
w_{min}&=&1+(t-1)(q-1),\\
w_{max }&=&t-1+\binom{t-1}{2}(q-2)+(t-1)(q-1).
\end{eqnarray*}
and then 
\begin{eqnarray*}
\frac{w_{min}}{w_{max }}&=&\frac{1+(t-1)(q-1)}{t-1+\binom{t-1}{2}(q-2)+(t-1)(q-1)}\\
&=&\frac{2}{t-1}\times \frac{1+(t-1)(q-1)}{(t-2)(q-2)+2q}.
\end{eqnarray*}
Now we prove that $\mathcal{C}$ is minimal. Suppose that $Supp(\omega)\subset Supp(\omega^{\prime})$ for some $\omega,\omega^{\prime}\in \mathcal{C}^*$ and that $\omega$ and $\omega^{\prime}$ are  linear combinations of $\emptyset\neq I_{\omega}\subset \{1,\ldots,t\}$ and $\emptyset\neq I_{\omega^{\prime}}\subset \{1,\ldots,t\}$ rows of $A_{t,q}$. Looking at the first $t$ coordinates of $\omega$ and $\omega^{\prime}$ one sees immediately that $I_{\omega}\subset I_{\omega^{\prime}}$. 

Consider now two rows $i\in  I_{\omega}$ and $j\in  I_{\omega^{\prime}}\setminus  I_{\omega}$. There exists precisely one entry corresponding to $e_i+\lambda e_j$ (or $e_j+\lambda e_i$) for which $\omega$ has a nonzero entry whereas  $\omega^{\prime}$ has a zero entry. This contradicts $Supp(\omega)\subset Supp(\omega^{\prime})$. So $I_{\omega}=I_{\omega^{\prime}}$. 

Therefore,
$$\omega=\alpha_1 r_{i_1}+\cdots+ \alpha_s r_{i_s}, \qquad \omega^{\prime}=\beta_1 r_{i_1}+\cdots +\beta_s r_{i_s},$$
for some $\alpha_i,\beta_i\in \mathbb{F}_q^*$, where $r_{i_j}$ denotes the $i_j$-th row of $A_{t,q}$. If $s=1$, there is nothing to prove since $\omega$ and $\omega^{\prime}$ are proportional. 

Suppose $s\geq 2$ and consider $1\leq n<m\leq s$ such that 
$$\alpha_n/\beta_n\neq \alpha_m/\beta_m.$$
Among the entries  corresponding to $e_i+\lambda e_j$, the unique zero in $\omega$ and $\omega^{\prime}$ appears respectively when $\alpha_n+\overline{\lambda}\alpha_m=0$ and $\beta_n+\overline{\mu}\beta_m=0$. Since $\alpha_n/\beta_n\neq \alpha_m/\beta_m$, $\overline{\lambda}\neq \overline{\mu}$ and therefore $Supp(\omega)\not \subset Supp(\omega^{\prime})$ and $Supp(\omega^{\prime})\not \subset Supp(\omega)$, a contradiction. 

This shows that 
$$\alpha_i/\beta_i= \alpha_j/\beta_j \qquad \forall i,j \in \{1,\ldots,s\}$$
and $\omega,\omega^{\prime}$ are proportional. Thus, $\mathcal{C}$ is minimal.
\endproof

\begin{remark}
Clearly, if $\mathcal{C}^{\prime}$ is an $\left[ n^{\prime}>n,t\right]_q$-code whose generator matrix $G^{\prime}$ contains $A_{t,q}$, then $\mathcal{C}^{\prime}$ is also minimal. 
\end{remark}

\section{Second construction}\label{Sec:Second_Construction}

Let $q$ be prime power. Fix two integers $t\geq 2$ and $2\le k\le t-1$ and consider the following matrix.

\begin{equation}\label{Eq:A_2}
\widetilde{A}_{t,q}=
\left(
\begin{array}{ccccc| ccccc}
&&&&&&&&&\\
&&&&&&&&&\\
&&\mathbb{I}_t&&&&&\mathbb{\widetilde{B}}_{t,q}&&\\
&&&&&&&&&\\
&&&&&&&&&\\
\end{array}
\right),
\end{equation}
where $\mathbb{I}_t$ is the identity matrix and $\mathbb{\widetilde{B}}_{t,q} \in \mathcal{M}_{t\times \binom{t}{k}(q-1)^{k-1}}(\mathbb{F}_q)$ is a matrix whose columns are $e_{i_1}+\displaystyle\sum_{j=2}^k\lambda_{i_j} e_{i_j}$, where $1\leq i_{j}<i_{l} \leq t$ for $1 \leq j<l \leq k$ and $\lambda_{i_j}\in \mathbb{F}_q^*$.

\begin{proposition}\label{Prop:Parameters_Secondo}
Let $q\ge t-2$. Then the linear code $\mathcal{D}$ whose generator matrix is $\widetilde{A}_{t,q}$ is a $\left[ \binom{t}{k}(q-1)^{k-1}+t,t,\tilde{d}\right]_q$-code, where $\tilde{d}\le 1+\binom{t-1}{k-1}(q-1)^{k-1}$.
\end{proposition}
\proof
Clearly, the length is $\binom{t}{k}(q-1)^{k-1}+t$ and the dimension is $t$. 
It is readily seen that the first row of the generator matrix has weight exactly
$$1+\binom{t-1}{k-1}(q-1)^{k-1}.$$
\endproof

\begin{proposition}\label{Prop:Second_Minimality}
The linear code $\mathcal{D}$ whose generator matrix is $\widetilde{A}_{t,q}$ is a minimal code.
\end{proposition}
\proof
Suppose that $\omega,\omega^{\prime}\in \mathcal{D}^*$ with $Supp(\omega)\subset Supp(\omega^{\prime})$ and that $\omega$ and $\omega^{\prime}$ are linear combinations of $\emptyset\neq I_{\omega}\subset \{1,\ldots,t\}$ and $\emptyset\neq I_{\omega^{\prime}}\subset \{1,\ldots,t\}$ rows of $\widetilde{A}_{t,q}$. Looking at the first $t$ coordinates of $\omega$ and $\omega^{\prime}$ one sees immediately that $I_{\omega}\subset I_{\omega^{\prime}}$.

Suppose that $I_{\omega}\neq I_{\omega^{\prime}}$ and consider the rows $ r_{\ell_1} $ and  $ r_{\ell_2} $, where $\ell_1\in  I_{\omega}$ and $\ell_2\in  I_{\omega^{\prime}}\setminus  I_{\omega}$. There exists at least one entry corresponding to one of the following:
\begin{enumerate}
    \item $e_{\ell_1}+\lambda_{\ell_2}e_{\ell_2}+\displaystyle\sum_{j=3}^k\lambda_{i_j} e_{i_j}$,
    \item $e_{\ell_2}+\lambda_{\ell_1}e_{\ell_1}+\displaystyle\sum_{j=3}^k\lambda_{i_j} e_{i_j}$,
    \item $e_{i_1}+\lambda_{\ell_1}e_{\ell_1}+\lambda_{\ell_2}e_{\ell_2}+\displaystyle\sum_{j=4}^k\lambda_{i_j}e_{i_j}$,
\end{enumerate}
for some $\lambda_{i_{j}}\in \mathbb{F}_q^*$, 
for which $\omega$ has a nonzero entry whereas $\omega^{\prime}$ has a zero entry. This contradicts $Supp(\omega)\subset Supp(\omega^{\prime})$. Thus, $I_{\omega}=I_{\omega^{\prime}}$.

If $|I_{\omega}|=1$, there is nothing to prove since $\omega$ and $\omega^{\prime}$ are proportional. From now on we consider the case $|I_{\omega}|=|I_{\omega^{\prime}}|=s\geq 2$ and  
\begin{eqnarray*}
\omega&=&(\alpha_1,\ldots,\alpha_t)\widetilde{A}_{t,q}\\
\omega^{\prime}&=&(\beta_1,\ldots,\beta_t)\widetilde{A}_{t,q},
\end{eqnarray*}
where $\alpha_{i}\neq 0$ and $\beta_i\neq 0$ if and only if $i\in I_{\omega}=I_{\omega^{\prime}}$. 

Consider a set $J=\{i_1,\ldots,i_k\}\subset \{1,\ldots,t\}$ such that $|J\cap I_{\omega}|\geq 2$.

 The entries corresponding to $\displaystyle e_{i_1}+\sum_{j=2}^k\lambda_{i_j} e_{i_j}$ are $$\alpha_{i_1}+\sum_{j=2}^{k}\lambda_j \alpha_{i_j}~\textrm{ and }~ \beta_{i_1}+\sum_{j=2}^{k}\lambda_j \beta_{i_j},$$
    respectively. Since the vanishing entries of $\omega$ and $\omega^{\prime}$ are at the same positions, the solutions (in $\lambda_i$) of 
    $\displaystyle \alpha_{i_1}+\sum_{j=2}^{k}\lambda_j \alpha_{i_j}=0$ and 
    $\displaystyle\beta_{i_1}+\sum_{j=2}^{k}\lambda_j \beta_{i_j}=0$ are the same. This happens only if $(\alpha_{i})_{i\in J\cap I_{\omega}}$ and $(\beta_{i})_{i\in J\cap I_{\omega}}$ are proportional.
    
    Since the above argument holds for any choice of the set $J$, 
    $(\alpha_{i})_{i\in I_{\omega}}$ and $(\beta_{i})_{i\in  I_{\omega}}$ are proportional, that is $\omega$ and $\omega^{\prime}$ are proportional. Thus, $\mathcal{D}$ is minimal.
\endproof

The weight distribution of $\mathcal{D}$ seems hard to be computed. 

\begin{problem}
	Determine the weight distribution of $\mathcal{D}$.
\end{problem}

Now we investigate another  class of minimal codes and we determine its weight distribution. 

\begin{theorem}
Let $\mathcal{D}^\prime$ be the code generated by the matrix $$\overline{A}_{s,t}=(\mathbb{I}_t|\overline{B}_{s,t}),$$ where $\overline{B}_{s,t}$ is a matrix whose columns are all the possible vectors of $\mathbb{F}_q^t$ of weight $s\leq t$. Let $N=t+\binom{t}{s}(q-1)^s$.
Denote by $\psi(r)$
$$\sum_{z=2}^r \binom{r}{z}\binom{t-r}{s-z}(q-1)^{s-z} \sum_{i=1}^{z-1}(q-1)^i(-1)^{z-1+i}.$$
Then $\mathcal{D}^\prime$ is minimal and the weight distribution of $\mathcal{D}^\prime$ is 

$$\left\{
N-t+r-\psi(r)-	\binom{t-r}{s}(q-1)^s :\
 r=0,\dots, s\right\}.$$
\end{theorem}
\proof
First note that $\mathcal{D}^\prime$ is minimal since  $\overline{A}_{s,t}$   contains  $\widetilde{A}_{t,q}$.

To determine the weight distribution, consider a codeword $w$ of type $(\alpha_1,\ldots,\alpha_t)\overline{A}_{s,t}$ with $(\alpha_1,\ldots,\alpha_t)$ of weight $r$. Among the first $t$ coordinates of $w$, exactly $r$ are nonzero.

Consider now all the coordinates $\mathcal{J}$ of $w$ corresponding to vectors in $\langle e_{i_1},\ldots,e_{i_s}\rangle$. These vectors are in total $(q-1)^s$. 

Let $z$ be the size of 
$$\mathcal{I}=\{i_j\ |\ \alpha_{i_j}\neq 0.\}$$

If $z=0$, then for each $i_1,\ldots,i_s$ we have that $\alpha_{i_j}=0$ and all the coordinates of  $w$ in $\mathcal{J}$ are $0$. There are precisely $\binom{t-r}{s}$ choices for $\{i_1,\ldots,i_s\}$ in this case.

Consider now the case $z>0$. The number of zero coordinates in $\mathcal{J}$ is $(q-1)^{s-z}\displaystyle\sum_{i=1}^{z-1}(q-1)^i(-1)^{z-1+i}$. 
In fact, these zeros  correspond to the number of solutions of the linear homogeneous equation
$$\sum_{i \notin \mathcal{I}}  0 X_i+ \sum_{i \in \mathcal{I}}  \alpha_i X_i=0$$
not satisfying any $X_i=0$. 

(Clearly, $z=1$ gives 0 for $\sum_{i=1}^{z-1}(q-1)^i(-1)^{z-1+i}$) 

Note that for a fixed $z>0$, then number of possibilities for $\{i_1,\ldots,i_s\}$ is $\binom{r}{z}\binom{t-r}{s-z}$. 

Summing up, we have that the number of zero coordinates is 

$$t-r+\psi(r)	+	\binom{t-r}{s}(q-1)^s.$$
and the claim follows.
\endproof

\begin{remark}
Suppose $s^2\leq 3t$. Then $\binom{r+1}{z}\binom{t-r-1}{s-z}\geq \binom{r}{z}\binom{t-r}{s-z}$ and so $\psi(r+1)>\psi(r)$. Therefore, the minimum weight of $\mathcal{D}^{\prime}$ is obtained for $r=s$.
\end{remark}

\section{A general construction}
\begin{theorem}\label{Th:General}
Let $\mathcal{C}$ be an $[n,k]_q$ minimal code with generator matrix $G$ such that 
\begin{equation}\label{Propr1}
\forall w=(w_1,\ldots,w_n) \in \mathcal{C}^*\ \Longrightarrow \ |\{w_i \ : \ i \in \{1,\ldots,n\}\}|=q.
\end{equation}
Then for any integer $s\geq1$ there exists an $[(s+1)n,s+k]_q$ minimal code $\mathcal{D}$ satisfying Property \eqref{Propr1}.  
\end{theorem}
\proof
Without loss of generality we can suppose that  $G$ is of the type $(\mathbb{I}_k|A)$ for some $A\in \mathcal{M}_{k\times n}(\mathbb{F}_q)$. Consider the code $\mathcal{D}$ whose generator matrix is $\overline{G}=(G_0|G_1|\cdots|G_s)$, where 

\begin{equation}\label{Eq:G_0}
G_0=
\left(
\begin{array}{ccccc}
0&0&\cdots&0&0\\
0&0&\cdots&0&0\\
\vdots&\vdots& &\vdots&\vdots\\
0&0&\cdots&0&0\\
0&0&\cdots&0&0\\
\hline
\\
&&G&&	\\
\\
\end{array}
\right) \in \mathcal{M}_{{(s+k)\times n}}(\mathbb{F}_q),
\end{equation}
\begin{equation}\label{Eq:G_i}
G_i=
\left(
\begin{array}{ccccc}
0&0&\cdots&0&0\\
\vdots&\vdots& &\vdots&\vdots\\
0&0&\cdots&0&0\\
1&1&\cdots&1&1\\
0&0&\cdots&0&0\\
\vdots&\vdots& &\vdots&\vdots\\
0&0&\cdots&0&0\\
\hline
\\
&&G&&	\\
\\
\end{array}
\right) \in \mathcal{M}_{{(s+k)\times n}}(\mathbb{F}_q),
\end{equation}
that is the $i$-th row is $\overline{1}$ and the remaining of the first $s$ rows of $G_i$ are $\overline{0}$. 
The dimension and the length of $\mathcal{D}$ are  $s+k$ and~$(s+1)n$, respectively. It is straightforward to check that $\mathcal{D}$ satisfies Property \eqref{Propr1}. 

We prove now that $\mathcal{D}$ is a minimal code. 
For a codeword 
\begin{eqnarray*}
\omega&=&(\omega_0,\omega_1,\ldots,\omega_s)\in \mathcal{D},
\end{eqnarray*} 
denote by $\omega_j$, $j=0,\ldots, s$, the vector 
$$\omega_j=(a_{j,1},\ldots, a_{j,n}).$$
Consider another codeword 
\begin{eqnarray*}
\omega^{\prime}&=&(\omega_0^{\prime},\omega_1^{\prime},\ldots,\omega_s^{\prime})
\end{eqnarray*} 
with $\omega_j^{\prime}=(b_{j,1},\ldots, b_{j,n})$, $j=0,\ldots,s$, 
such that $Supp(\omega)\subset Supp(\omega^{\prime})$. 
Also, let 
\begin{eqnarray*}
\omega&=&\sum_{i=1}^s \alpha_i R_i+\sum_{i=s+1}^{s+k} \beta_i R_i,\\
\omega^{\prime}&=&\sum_{i=1}^s \alpha_i^{\prime} R_i+\sum_{i=s+1}^{s+k} \beta_i^{\prime} R_i,\\
\end{eqnarray*}
where $R_i$ denotes the $i$-th row of $\overline{G}$. Since $Supp(\omega_0)\subset Supp(\omega^{\prime}_0)$ and $\mathcal{C}$ is minimal, there exists $\mu\in \mathbb{F}_q^*$ such that $\mu \beta_i = \beta_i^{\prime} $ for each $i=s+1,\ldots, s+k$. 

Since $\mathcal{C}$ satisfies Property \eqref{Propr1}, there are $q$ distinct coordinates in $\omega_0$.  Let $i_1,\ldots,i_q\in \{1,\ldots,n\}$ be such that 
$$|\{a_{0,i_\ell} \ : \ 1\leq \ell\leq q\}| =q.$$
Now for  $i=1,\ldots,s$, we consider
\begin{eqnarray*}
Supp(\omega_0+\alpha_i\overline{1})&=& Supp(\omega_i)\subset Supp(\omega^{\prime}_i)\\
&=&Supp(\omega_0^{\prime}+\alpha_i^{\prime}\overline{1})\\
&=&Supp(\mu \omega_0+\alpha_i^{\prime}\overline{1}).
\end{eqnarray*}
In particular, there exists a unique $\overline{\ell} \in \{1,\ldots,q\}$ such that $a_{0,i_{\overline{\ell}}}+\alpha_i=0$. Since 
$$
|\{a_{0,i_\ell}+\alpha_i \ : \ 1\leq \ell\leq q\}| = q
= |\{\mu a_{0,i_\ell}+\alpha_i^{\prime} \ : \ 1\leq \ell\leq q\}|,
$$
the unique zero among $\mu a_{0,i_\ell}+\alpha_i^{\prime}$, $\ell=1,\ldots,q$, must be at the same position as $a_{0,i_{\overline{\ell}}}+\alpha_i$. That is, 
$\mu a_{0,i_{\overline{\ell}}}+\alpha_i^{\prime}=0$ and therefore $\alpha_i^{\prime}=-\mu a_{0,i_{\overline{\ell}}}=\mu \alpha_i$. This means that $\omega^{\prime}=\mu \omega$ and $\mathcal{D}$ is minimal.
\endproof

\begin{corollary}
Let $q$ be prime power. 
Consider $\mathbb{F}_{q}^*=\langle \xi\rangle$. 
Fix an integer $t\geq 2$ and consider the following matrix.
\begin{equation}\label{Eq:Aprime}
A^{\prime}_{t,q}=
\left(
\begin{array}{ccccc|  ccccc}
&&&&&\xi&\xi^2&\cdots&\xi^{q-3}&\xi^{q-2}\\
&&&&&0&0&\cdots&0&0\\
&&A_{t,q}&&&0&0&\cdots&0&0\\
&&&&&0&0&\cdots&0&0\\
&&&&&0&0&\cdots&0&0\\
\end{array}
\right),
\end{equation}
where $A_{t,q}$ is the  matrix defined in \eqref{Eq:A}. The code $\mathcal{C}^{\prime}$ generated by $A^{\prime}_{t,q}$ is a minimal $\left[ \binom{t}{2}(q-1)+t+q-2,t\right]_q$-code satisfying Property \eqref{Propr1}. 
\end{corollary}
\proof
Since the code $\mathcal{C}$ generated by $A_{t,q}$ is minimal and it is a subcode of $\mathcal{C}^{\prime}$, this is also minimal. We only have to check that Property \eqref{Propr1} is satisfied. Let us denote by $R_i$, $i=1,\ldots,t$, the rows of $A^{\prime}_{t,q}$.

\begin{itemize}
\item Let $\omega=\alpha_1 R_1$. Then 
$$\omega=(\alpha_1,0,\ldots,0,\xi\alpha_1 ,\xi^2\alpha_1 ,\ldots,\xi^{q-3}\alpha_1 ,\xi^{q-2}\alpha_1)$$ and Property \eqref{Propr1} holds for $\omega$. 
\item Let $\omega=\alpha_i R_i$, $i>1$. Then the entries corresponding to $e_1+\lambda e_i$ are $\lambda  \alpha_i $ and, therefore, they are all distinct and nonzero. 
\item Let $\omega=\sum_{i}\alpha_i R_i$, where at least two $\alpha_i$'s are nonzero, say $\alpha_{\ell}$ and $\alpha_j$. Then the entries corresponding to $e_\ell+\lambda e_j$ are $\alpha_\ell+\lambda \alpha_j$. Combined with the $\ell$-th entry of $\omega$ (which is $\alpha_\ell$) all such entries are distinct and Property \eqref{Propr1} holds for $\omega$.
\end{itemize}
The claim follows from Theorem \ref{Th:General}. Let $\omega$ be a codeword of $C^{\prime}$. If  it is spanned from $s$ rows of $A^\prime_{t,q}$ distinct from the first one, its weight is  
$$
w(\omega)=w_s=s+\binom{s}{2}(q-2)+s(t-s)(q-1)
$$
for $1\le s\le t-1$. On the other hand, if $\omega$ is the linear combination of the first row and other $(s-1)$ rows, its weight is 
$$
w(\omega)=w_s=s+\binom{s}{2}(q-2)+s(t-s)(q-1)+(q-1)
$$
for $1\le s\le t$.
\endproof

\begin{example}
\label{Example1}
	For $q$ odd, let $f:\mathbb{F}_q^n\to \mathbb{F}_q$ be the function introduced in \cite{BB2019}, that is,
	\[
	f(x) = 
	\begin{cases}
	\alpha_i, & {\rm wt}(x)=i\le k,\\
	0,& {\rm wt}(x)> k,\\
	\end{cases}
	\]
	where $n> 3$ is an integer, $k\in \{2,\ldots,n-2\}$, and  $\{\alpha_i\}_{i\in\{1,\ldots,k\}}$ are (not necessarily distinct) elements of $\mathbb{F}_q^*$.
	Let $\mathcal{C}_f$ be the linear code defined as
	\begin{equation}\label{Def:Code}
	\mathcal{C}_f :=\{(uf(x)+v\cdot x)_{x \in \mathbb{F}_q^n \setminus \{0\}} \ | \ u \in \mathbb{F}_q, v \in \mathbb{F}_q^n\},
	\end{equation}
	where $v\cdot x$ is the Euclidean inner product between $v$ and $x$. 
	
	For any pair $(u,v) \in \mathbb{F}_q\times \mathbb{F}_{q}^n$, let us denote 
	$$
	c(u,v):=(uf(x)+v\cdot x)_{x \in \mathbb{F}_q^n \setminus \{0\}}.
	$$
	
	If $k\ge q$ and $\{\alpha_i:\,i=1,\dots,k\}=\mathbb{F}_q$, then Property $\eqref{Propr1}$ is satisfied.
	
	\begin{itemize}
		\item if $u=0$, then $c(u,v)=(v\cdot x)_{x\in\mathbb{F}_q^n}$ has as components all the elements of $\mathbb{F}_q$ since the scalar product is a linear function;
		\item if $v=0$, then $c(u,v)=(uf(x))_{x\in\mathbb{F}_q^n}$ has as components all the elements of $\mathbb{F}_q$ (e.g. they come from the standard basis of $\mathbb{F}_q^n$ over $\mathbb{F}_q$);
		\item otherwise, we show that $c(u,v)=(uf(x)+v\cdot x)_{x\in\mathbb{F}_q^n}$ has as components all the elements of $\mathbb{F}_q$: take $i$ such that $v_i\ne0$ and consider the vectors $\beta e_i$, where $\beta\in\mathbb{F}_q$. The components of $c(u,v)$ corresponding to these components will be $u\alpha_i+\beta v_i$, which will arise all the possible values of $\mathbb{F}_q$.
	\end{itemize}
	\end{example}

	\begin{example}
	Let $n=rk$, where $r,k\in\mathbb{N}$ and $r,k\ge2$, be a positive integer and consider the function, introduced in \cite{BB201?}, $g_{r,k}:\mathbb{F}_q^n\to \mathbb{F}_q$ 
	
	$$g_{r,k}(x_1,\ldots,x_n):=\sum_{j=0}^{k-1} x_{jr+1}x_{jr+2}\cdots x_{jr+r}.
	$$
	The code $C_{g_{r,k}}$ defined  in Example \ref{Example1} satisfies Property \eqref{Propr1}. 
	
	\begin{itemize}
		\item if $u={\bf 0}$, then the codeword  $c(u,v)=(v\cdot x)_{x\in\mathbb{F}_q^n}$ has as components all the elements of $\mathbb{F}_q$ since the scalar product is a linear function.
		\item if $v={\bf 0}$, then in the codeword $c(u,v)=(uf(x))_{x\in\mathbb{F}_q^n}$, the entries corresponding to vectors $(x_1=1,\ldots,x_{r-1}=1,x_r\in \mathbb{F}_q,x_{r+1}=0,\ldots,x_n=0)$ assume all the possible values of $\mathbb{F}_q$.
		\item if $u\neq {\bf 0}$ and $ v\neq {\bf 0}$, let $i$ be such that $v_i\ne0$ and consider the vectors $\beta e_i$, where $\beta\in\mathbb{F}_q$. The components of $c(u,v)$ corresponding to these vectors are 
		$\{\beta v_i | \beta \in \mathbb{F}_q\}=\mathbb{F}_q$. 
	\end{itemize}
	\end{example}
\section*{Acknowledgment}

The research of D. Bartoli and M. Bonini was supported by the Italian National Group for Algebraic and Geometric Structures and their Applications (GNSAGA - INdAM).

\ifCLASSOPTIONcaptionsoff
  \newpage
\fi


\begin{thebibliography}{1}


\bibitem{AB1998}
\newblock A. Ashikhmin and A. Barg,
\newblock ``Minimal vectors in linear codes,"
\newblock \emph{IEEE Trans. Inf. Theory}, vol. 44, no. 5, pp. 2010--2017,  1998.

\bibitem{BB2019}
\newblock D. Bartoli and M. Bonini,
\newblock ``Minimal Linear Codes in Odd Characteristic,"
\newblock \emph{IEEE Trans. Inform. Theory},  vol. 65, no. 7, pp. 4152--4155, 2019. 

\bibitem{BMeT1978} 
\newblock E. R Berlekamp, R. J. McEliece, and H. C. A. van Tilborg,
\newblock ``On the Inherent Intractability of Certain Coding Problems,"
\newblock \emph{IEEE Trans. Inform. Theory}, vol. 24, no. 3, pp. 384--386, 1978.

\bibitem{Blakley1979} 
\newblock G. R. Blakley,
\newblock ``Safeguarding cryptographic keys,"
\newblock in \emph{Proc. of AFIPS National Computer Conference}, New York, USA, 1979, pp. 313--317.

\bibitem{BB201?} 
\newblock M. Bonini and M. Borello,
\newblock ``Minimal linear codes arising from blocking sets,"
\newblock \emph{Journal of Algebraic Combinatorics}, to appear (arXiv:1907.04626).


\bibitem{MAGMA} 
W. Bosma, J. Cannon, and C. Playoust,  
\newblock ``The Magma algebra system. I. The user language. Computational algebra and number theory," 
\newblock {\em J. Symbolic Comput.}, vol. 24, no. 3-4, pp. 235--265, 1997.


\bibitem{CDY2005} 
\newblock C. Carlet, C. Ding, and J. Yuan,
\newblock ``Linear codes from highly nonlinear functions and their secret sharing schemes,"
\newblock \emph{IEEE Trans. Inf. Theory}, vol. 51, no. 6, pp. 2089--2102, 2005.


\bibitem{CCP2014}
\newblock H. Chabanne, G. Cohen, and A. Patey,
\newblock ``Towards Secure Two-Party Computation from the Wire-Tap Channel," 
\newblock in \emph{Information Security and Cryptology -- ICISC 2013}, Heidelberg, Germany,  2014, pp. 34--46.

\bibitem{CH2017}
\newblock S. Chang and J. Y. Hyun,
\newblock ``Linear codes from simplicial complexes,"
\newblock \emph{Des. Codes Cryptogr.}, vol. 86, no. 10, pp. 2167--2181, 2018.

\bibitem{BN1990} 
\newblock J. Bruck and M. Naor,
\newblock ``The Hardness of Decoding Linear Codes with Preprocessing,"
\newblock \emph{IEEE Trans. Inform. Theory}, vol. 36, no. 2, pp. 381--385,  1990.

\bibitem{CMP2013} 
\newblock G. D. Cohen, S. Mesnager, and A. Patey,
\newblock ``On minimal and quasi-minimal linear codes,"
\newblock in \emph{IMACC 2013}, Heidelberg, Germany,  2013, pp. 85--98.


\bibitem{Ding2015} 
\newblock C. Ding,
\newblock ``Linear codes from some $2$-designs,"
\newblock \emph{IEEE Trans. Inf. Theory}, vol. 60, no. 6, pp. 3265--3275,  2015.


\bibitem{DHZ2018}
\newblock C. Ding, Z. Heng, and Z. Zhou,
\newblock ``Minimal binary linear codes,"
\newblock \emph{IEEE Trans. Inf. Theory}, vol. 64, no. 10, pp. 6536--6545, 2018.


\bibitem{DLLZ2016}
\newblock C. Ding, N. Li, C. Li, and Z. Zhou,
\newblock ``Three-weight cyclic codes and their weight distributions,"
\newblock \emph{Discrete Math.}, vol. 339, no. 2, pp. 415--427,  2016.


\bibitem{DY2003} C. Ding and J. Yuan,
\newblock ``Covering and secret sharing with linear codes,"
\newblock in \emph{DMTCS 2003}, Heidelberg, Germany, 2003, pp. 11--25.


\bibitem{HDZ2018}
\newblock Z. Heng, C. Ding, and Z. Zhou,
\newblock ``Minimal Linear Codes over Finite Fields,"
\newblock \emph{Finite Fields Appl.}, vol. 54, 2018, pp. 176-196.


\bibitem{HirschBook}
\newblock J. W. P. Hirschfeld,
\newblock \emph{Projective geometries over finite fields.} 2nd edition,
\newblock Oxford Univ. Press, Oxford, England, 1998.

\bibitem{Klove}  Klove, T.: Codes for error detection.  Series on Coding Theory and Cryptology, 2. World Scientific Publishing Co. Pte. Ltd., Hackensack, NJ, 2007.


\bibitem{Massey1993}
\newblock J. L. Massey,
\newblock ``Minimal codewords and secret sharing,"
\newblock in \emph{Proc. 6th Joint Swedish-Russian Int. Workshop on Info. Theory}, M\"olle, Sweden,   1993, pp. 276--279.

\bibitem{Massey1995}
\newblock J. L. Massey,
\newblock ``Some applications of coding theory in cryptography,"
\newblock in \emph{Codes and Cyphers: Cryptography and Coding IV}, Esses, England, 1995, pp. 33--47.

\bibitem{Shamir1979} 
\newblock A. Shamir,
\newblock ``How to share a secret,"
\newblock \emph{Commun. ACM}, vol. 22, no. 11, pp 612--613,  1979.


\bibitem{SL2012} 
\newblock Y. Song and Z. Li,
\newblock ``Secret sharing with a class of minimal linear codes," available online at:
\newblock https://arxiv.org/abs/1202.4058,  2012.


\bibitem{YD2006}
\newblock J. Yuan and C. Ding,
Secret sharing schemes from three classes of linear codes,"
\newblock \emph{IEEE Trans. Inf. Theory}, vol. 52, no. 1, pp. 206--212, 2006.


\end{thebibliography}
\end{document}